\documentclass[traditabstract]{aa}
\pdfoutput=1
\usepackage[pdftex]{color,xcolor}
\usepackage[pdftex]{graphicx}
\usepackage{hyperref}               
\hypersetup{pdfauthor=Christoph Mordasini}

\usepackage{txfonts}
\usepackage{wasysym}
\usepackage{marvosym}
\def\mearth{M_\oplus}
\def\msun{M_\odot}

\def\mnept{M_{\textrm{\tiny{\neptune}}}}

\def\msini{M\sin i}
\def\harps{{\footnotesize HARPS} }
\def\lsun{L_{\odot}}
\def\msun{M_{\odot}}

\def\aj{AJ}                   
\def\araa{ARA\&A}             
\def\apj{ApJ}                 
\def\apjl{ApJ}                
\def\apjs{ApJS}               
\def\ao{Appl.Optics}          
\def\aap{A\&A}                
\def\mnras{MNRAS}             
\def\pasp{PASP}               

\begin{document}

\title{The HARPS search for southern extra-solar planets 
\thanks{Based on observations made with the {\footnotesize HARPS} instrument on the ESO 3.6 m telescope at La Silla Observatory under the GTO programme ID 072.C-0488.}\fnmsep
\thanks{Individual radial velocities are available electronically  at  CDS via anonymous \texttt{ftp} to \texttt{cdsarc.u-strasbg.fr (130.79.128.5)} or via \texttt{http://cdsweb.u-strasbg.fr/cgi-bin/qcat?J/A+A/}}
}

\subtitle{XXIV. Companions to HD\,85390, HD\,90156 and HD\,103197:\\ A Neptune analogue and two intermediate mass planets}

 \author{ C.~Mordasini \inst{1,2} 
       \and M.~Mayor\inst{3} 
        \and S.~Udry\inst{3}
        \and C.~Lovis\inst{3} 
        \and D.~S\'egransan\inst{3}  
        \and W.~Benz\inst{2} 	  
        \and J.-L.~Bertaux\inst{4}
        \and F.~Bouchy\inst{5}    
        \and G.~Lo Curto \inst{6}
        \and C.~Moutou\inst{7}  
        \and D.~Naef \inst{3,6}   
        \and F.~Pepe\inst{3} 
        \and D.~Queloz\inst{3} 
        \and N.~C.~Santos \inst{8}	
      } 

   \institute{
               Max-Planck-Institut f\"ur Astronomie, 
               K\"onigstuhl 17, 69117 Heidelberg, Germany \\
               \email{mordasini@mpia.de}
            \and
             Physikalisches Institut, Universit\"at Bern, Sidlerstrasse 5, 3012 Bern, Switzerland
  	   \and 
              Observatoire de Gen\`eve, Universit\'e  de Gen\`eve
              51 Ch. des Maillettes, 1290  Sauverny, Switzerland
          \and
             Service d'A\'eronomie du CNRS/IPSL, Universit\'e de Versailles Saint-Quentin, BP 3, 91371 Verri\`eres-le-Buisson, France
          \and
             Institut d'Astrophysique de Paris, CNRS, Universit\'e
             Pierre et Marie Curie, 98bis Bd Arago, 75014 Paris,
             France
               \and
             European Southern Observatory, Casilla 19001, Santiago 19, Chile 
          \and      
	        Laboratoire d'Astrophysique de Marseille, Traverse du Siphon, 13376 Marseille 12, France
          \and
             Centro de Astrof\'{i}sica da Universidade do Porto, Rua das Estrelas, 4150-762 Porto, Portugal
            	     }

\date{Received / Accepted  To be inserted later}
\abstract{We report the detection of three new extrasolar planets orbiting the solar type stars \object{HD\,85390}, \object{HD\,90156} and \object{HD\,103197} with the {\footnotesize HARPS} spectrograph  mounted on the ESO 3.6-m telescope at La Silla observatory. \object{HD\,85390} has a planetary companion with a projected intermediate mass  ($42.0\, \mearth$) on a 788-day orbit ($a$=1.52 AU) with an eccentricity of 0.41, for which there is no analogue in the solar system. A drift in the data indicates the presence of another companion on a long period orbit, which is however not covered by our measurements. \object{HD\,90156} is orbited by a warm Neptune analogue with a minimum mass of 17.98 $\mearth$ (1.05 $\mnept$), a period of  49.8 days ($a$=0.25 AU) and an eccentricity of 0.31.  \object{HD\,103197} has an intermediate mass planet on a circular orbit ($P=47.8$ d, $\msini=31.2\, \mearth$). We discuss the formation of planets of intermediate mass ($\sim30-100\, \mearth$) which  should be rare inside a few AU according to core accretion formation models.}


\titlerunning{The  {\footnotesize HARPS} search for southern extra-solar planets. XXIV.}
\authorrunning{C. Mordasini et al.}

\maketitle

\section{Introduction}\label{sect:intro}
The \harps (High Accuracy Radial Velocity Planet Searcher) program has been on-going since 2003 at the ESO 3.6 meter telescope located at La Silla Observatory in Chile (Pepe \cite{pepeetal2002}). Thanks to its high accuracy of $\lesssim 1$ m/s and its long time stability, \harps has in particular detected many low mass planets, starting with $\mu$ Arae b at the beginning of the program  (Santos et al. \cite{santosetal2004}) to the recent two Earth-mass planet around \object{GJ 581} (Mayor et al. \cite{mayoretal2009}). As such low mass planets seem to be very abundant (Lovis et al. \cite{lovisetal2009}), \harps helps significantly to increase the number of known extrasolar planets. The latter is raising rapidly in time $t$, roughly as $t^{2..3}$ since the discovery of 51 Peg b by Mayor \& Queloz (\cite{mayorqueloz1995}) fifteen  years ago. 

These detections of extrasolar planets have enormously stimulated the research on planet formation. They have shown the importance of mechanisms which were previously underestimated in their importance from the shape of our own solar system alone, like for example orbital migration which was known on purely theoretical bases for along time (Goldreich \& Tremaine, \cite{goldreichtremaine1980}), overthrowing the idea that giant planets can only be found beyond a few astronomical units (Boss \cite{boss1995}). 

The quickly growing number of extrasolar planets has allowed to derive many important distributions of physical properties and orbital elements. We know now distributions of masses, semi-major axes or eccentricities of giant planets within a few AU around solar like stars (e.g. Udry \& Santos \cite{udrysantos2007}), and see correlations between stellar and planetary parameters. The most important example is the ``metallicity effect'' i.e. the increase of giant planet frequency with stellar [Fe/H], see e.g. Santos et al. (\cite{santosetal2001}) or Fischer \& Valenti (\cite{fischervalenti2005}).  

By varying initial conditions of planetary formation models in a Monte Carlo way (Ida \& Lin \cite{idalin2004}), one can try to derive these statistical distributions also from theoretical formation models and compare them with the observed ones. This young method of planetary population synthesis (Mordasini et al. \cite{mordasinietal2009a}, \cite{mordasinietal2009b}) allows to use the full wealth of observational data to constrain theoretical models and to improve  our understanding of planet formation, as demonstrated here for the specific case of intermediate mass planets. 

In this paper we report the discovery of  two types of planets, namely of one Neptunian planet and of two intermediate mass planets (about 31 and 42 $\mearth$) without analogue in the solar system. We discuss the formation of the latter objects in the context of the core accretion formation model at the end of the paper. Such intermediate mass planets are interesting, as they provide observational constraints on the rate at which gas can be accreted by planets starting runaway gas accretion.

The paper is organized as follows: In section \ref{sec:hoststarcharacteristics} we  discuss the host star properties. Radial velocity measurements and orbital solutions are presented in section \ref{sec:measurements}. In section \ref{sec:summary}, we summarize our discoveries and in section \ref{sec:discussion} we discuss them.

\section{Host star characteristics}\label{sec:hoststarcharacteristics} 
Basic photometric and astrometric properties for the three host stars HD\,85390 (\object{HIP\,48235}), HD\,90156 (\object{HIP\,50921}) and HD\,103197 (\object{HIP\,57931}) are taken from the Hipparcos catalogue (ESA 1997). Accurate spectroscopic stellar parameters and derived quantities for HD\,85390 and HD\,90156 are taken from the analysis of  Sousa et al. (\cite{sousaetal2008}). Sousa et al. (\cite{sousaetal2008}) used high quality, high signal to noise \harps spectra to obtain homogenous estimates of the spectroscopic stellar parameters of all 451 targets in the \harps Guaranteed Time Observations (GTO) ``high precision'' sample, to which the two mentioned stars belong.  HD\,103197 is in contrast part of the lower RV precision ``{\footnotesize CORALIE} extension'' sample, see Naef et al. (\cite{naefetal2007}).  For this star, we stacked several individual spectra together to obtain a high signal to noise spectrum and repeated the analysis described in Sousa et al. (\cite{sousaetal2008}). 

\begin{table}
\caption{Observed and inferred stellar parameters for the three host stars. Photometric and astrometric quantities were taken from the Hipparcos catalogue (ESA 1997) while stellar physical quantities come from Sousa et al. (\cite{sousaetal2008}), except for HD\,103197. For this star, the values are from the present paper, following the analysis in Sousa et al. (\cite{sousaetal2008}).}\label{tab:starprops}
\begin{center}
\begin{tabular}{llccc}
\hline\hline
Parameter                         & Unit           & HD\,85390   & HD\,90156           &HD\,103197           \  \\ \hline                                                        
Spectral type                       &             	     & K1V 		&  G5V               &     K1Vp   	         \\ 
V                                        & [mag]           & 8.54      	&   6.92              &       9.40    	      \\
$B-V$                                & [mag]           & 0.86		& 0.66               &        0.86  	 \\
$\pi$                                 & [mas]          &29.45$\pm$0.84 &45.26$\pm$0.75& 20.27$\pm$1.47 	 \\
$M_{V}$                           & [mag]           &5.89		&  5.20                   	    &  5.77          \\
$T_{\rm eff}$ 		        & [K]                &5186$\pm$54&   5599$\pm$12   & 5303$\pm$58 \\
$\log{g}$                          & [cgs]              &4.41$\pm$0.09& 4.48$\pm$0.02   & 4.40$\pm$0.11    \\
$\mathrm{[Fe/H]}$  		                & [dex]             &-0.07$\pm$0.03& -0.24$\pm$0.01  & 0.21$\pm0.04$    \\
$L$                                   &[L$_{\odot}$] & 0.43$\pm$0.03&  0.72$\pm$0.01   &  0.47$\pm$0.11         \\
$M_{\star}$             &[M$_{\odot}$]       & 0.76                    & 0.84       &     0.90                 \\
$v \sin{i}$                     & [km/s]   	  &1			&  $<$1      &  2            \\
$\log{R^{'}_{HK}}$   &           		 & -4.97			&   -4.95      &  -5.07        \\
$P_{\rm rot}$           &[days]     &44$\pm$5 & 26$\pm$3    & 51$\pm$5   \\
Age                            & [Gyr]      &$\sim7.2$ & $\sim4.4$    & $\sim9.1$ \\
\hline
\end{tabular}
\end{center}
\end{table}

Individual \harps spectra were used to derive the radial velocity RV, the Bisector Inverse Slope of the Cross-Correlation Function, the $v\sin{i}$  of the star as well as the chromospheric activity index $\log{R^{'}_{HK}}$, using a similar recipe as Santos et al. (\cite{santosetal2000}). From the activity indicator we also derive an estimate of the stellar rotation period $P_{\rm rot}$ using the empirical Noyes et al. (\cite{noyesetal1984}) activity-rotation correlation. Approximate ages are obtained using the improved activity-rotation-age calibration of Mamajek \& Hillenbrand (\cite{mamajekhillenbrand2008}). These ages typically have errors of about 0.2 dex.

The basic properties of the four stars are summarized in Table \ref{tab:starprops}. The stellar mass estimates from Sousa et al. (\cite{sousaetal2008}) typically have an error of 0.1 $\msun$.

\subsection{HD\,85390 (HIP\,48235)}
HD\,85390 is an K1 dwarf in the southern hemisphere. The Hipparcos catalog (ESA 1997) lists a visual magnitude $V=8.54$, a color index $B-V=0.86$, and an astrometric parallax $\pi=29.45\pm0.84$ mas, setting the star at a distance of $34\pm 1$ pc from the Sun, a typical distance for planet host stars in RV search programs.  Its absolute magnitude is then estimated to be $M_{V} = 5.89$ mag. The spectral analysis of Sousa et al. (\cite{sousaetal2008}) yields the following physical parameters: An effective temperature $T_{\rm eff}=5186\pm54$ K, a surface gravity $\log{g}=4.41\pm0.09$, and a metallicity [Fe/H]$=-0.07\pm0.03$ dex. The star has thus a metallicity similar to the Sun. The derived mass is $0.76\, \msun$, and the luminosity is estimated from the absolute magnitude, the effective temperature and the corresponding bolometric correction to be 0.43 $\lsun$, thus less than half  the solar luminosity.  A  low projected rotational velocity $v \sin{i}$ of about 1  km/s is derived from a calibration of the width of the cross-correlation function  as described in Santos et al. (\cite{santosetal2002}). With an activity indicator $\log{R^{'}_{HK}}$ of -4.97, the star is among the non-active stars in our sample, and no large radial velocity jitter is expected. The detailled way how we deal with the stellar jitter is described in section \ref{sec:jitter}.

Several additional measurements of the properties of HD\,85390 can be found in the literature that agree fairly well with the mentioned values: Gray et al. (\cite{grayetal2006}) list a spectral type K1.5V, $T_{\rm eff}=5069$ K, $\log{g}=4.48$, [M/H]$=-0.11$ and $\log{R^{'}_{HK}}$ of -5.06. Henry et al. (\cite{henryetal1996}) measure a $\log{R^{'}_{HK}}$ of -4.93. Minniti  et al. (\cite{minnitietal2009}) classify the star as a K1 radial velocity stable star at a precision of about 5 m/s.  

\subsection{HD\,90156 (HIP\,50921) } 
According to the Hipparcos catalogue (ESA 1997) is HD\,90156 a quite close G5V star with an astrometric parallax of $\pi=45.26\pm 0.75$\,mas with basic photometric properties of $V=6.92$  and $B-V$=0.66. Sousa et al. (\cite{sousaetal2008}) derived from the \harps spectra a $T_{\rm eff}=5599\pm12$ K, $\log{g}=4.48\pm0.02$, and a  subsolar metallicity of [Fe/H]$=-0.24\pm0.01$  dex which is well known to play an important role in planet formation. Other derived quantities are  $M_{\star}=0.84\,\msun$ and $L=0.72\pm0.01\, \lsun$. The star has a low projected rotation velocity  of less than 1 km/s which is thus difficult to determine accurately, see Melo et al. (\cite{meloetal2001})  and a low activity index of $\log{R^{'}_{HK}}=-4.95$.  

The star has been studied by several other groups: It was in particular spectroscopically analyzed by Valenti \& Fischer (\cite{valentifischer2005}).  By fitting synthetic spectra they find an effective temperature of $5626\pm44$ K and $\log{g}=4.63\pm0.06$.  This surface gravity is significantly larger than our value or their value calculated from isochrone interpolation ($\log{g}=4.46$), corresponding to a systematic discrepancy noted by the authors themselves.  The mass derived from their spectroscopic gravities is with $1.25\pm0.17\,\msun$ correspondingly considerably larger than our value. On the other hand is the mass they derive from interpolating isochrones with $0.90\pm0.04\, \msun$ similar to our result. Valenti \& Fischer (\cite{valentifischer2005}) also measure a subsolar metal content of [Fe/H]=$-0.21\pm0.03$ and estimate the age of the star to be 7.8 Gyr.  

The metal poor nature of the star is further confirmed by Neves et al. (\cite{nevesetal2009}) who note that the star is a member of the thin disk and by Gray et al. (\cite{grayetal2006}) who find a [M/H] of -0.26. The latter authors also give a 
$T_{\rm eff}=5578$ K,  $\log{g}=4.52$ and $\log{R^{'}_{HK}}=-4.96$ which is comparable to our values. HD\,90156 is  listed in the catalogue of suspected variable stars of Kukarkin et al. (\cite{kukarkinetal1981}) with an amplitude in V of 0.1 mag. The much more accurate Hipparcos data does however not point to any significant variability.  

\subsection{HD\,103197 (HIP\,57931)}
HD\,103197 is classified in the Hipparcos catalogue (ESA 1997) as K1Vp, with V=9.40, B-V=0.86 and $\pi=20.27\pm1.47$ mas. Repeating the same spectroscopic analysis as described in Sousa et al. (\cite{sousaetal2008}) with stacked \harps spectra leads to the following values: $T_{\rm eff}=5303\pm58$ K, $\log{g}$= $4.40\pm0.11$  and a supersolar metallicity of [Fe/H]$=0.21\pm0.04$  dex. The derived mass and luminosity is 0.90 $\msun$ and  0.47 $\lsun$. From the \harps spectra we also infer the following quantities:  a projected rotational velocity of roughly $v\sin{i}\approx 2$ km/s, a low activity indicator $\log{R^{'}_{HK}}=-5.07$ and an estimated $P_{\rm rot}$ of about $51\pm5$ days. This star is therefore very inactive. 

For comparison, Jenkins et al. (\cite{jenkinsetal2008}) find also a low $\log{R^{'}_{HK}}=-5.05$, but their  [Fe/H]=$-0.12\pm0.04$ is clearly lower than our value.
 
\section{Radial velocity measurements  and orbital solutions}\label{sec:measurements}  

\subsection{Jitter and error estimates} \label{sec:jitter}
The quantified error of an individual radial velocity measurement includes photon noise, calibration and instrumental drift uncertainty.  Additional, not quantified errors which are not included in this number will originate from intrinsic radial velocity variability of the star (jitter) due to granulation (Kjeldsen et al. \cite{kjeldsenetal2005}), magnetic activity and pulsations (see Pepe \& Lovis \cite{pepelovis2008} for an overview). This jitter can in principle be a serious threat to high precision RV measurement at a level of 1 m/s (Wright \cite{wright2005}). On the other hand, there are now severals stars in the \harps high precision program like for example \object{HD\,69830} (Lovis et al. \cite{lovisetal2006}) or \object{HD\,40307} (Mayor et al. \cite{mayoretal2009}), for which residuals around the fit clearly below 1 m/s have been found, like 0.64 m/s for HD\,69830 (Lovis et al. \cite{lovisetal2006}).

Neglecting the RV jitter in the total error estimate can have negative effects in the derivation of the uncertainties of the orbital parameters of the fits. Neglecting it has also the effect that measurements with a small quantified error (particularly high S/N, due to e.g. very good seeing to which \harps is sensitive) get an inappropriately high weight in the fitting procedure, causing also misleading results.   

Several authors have tried to derived empirical relations linking a star's radial velocity jitter to other quantities like $B-V$ or $R^{'}_{HK}$ (e.g. Saar et al. \cite{saaretal1998}; Wright et al. \cite{wright2005}). Concerning the latter work, all three stars presented here would fall in the class with the lowest derived median jitter of 3.5 m/s. The jitter of the stars presented here is however obviously significantly smaller, as can be seen by the fact that planets which induce a radial velocity semi-amplitude of that order are clearly detected in the data. This indicates that those older jitter estimates were clearly affected by instrumental and data reduction uncertainties.

We find from our \harps measurements that for the typical  stars in the high precision program with insignificant evolution, a very low activity level ($\log{R^{'}_{HK}}\leq-4.95$ like for the three stars in this work), a small intervall in $B-V$ (usually 0.5-1.0, 0.66-0.86 in this paper), and a small  $v\sin{i}$ ($\leq 2$ km/s), it is difficult to derive empirical relations like the ones mentioned before which cover much larger domains of these quantities. For example, we do not see an obvious correlation between the minimal RV jitter and $\log{R^{'}_{HK}}$ or $B-V$ among the stars with the mentioned properties in our sample. In a recent study, Isaacson \& Fischer (\cite{isaacsonfischer2010}) found at least qualitatively very similar conclusions, with in particular a jitter level independent of activity indicators and below their instrumental and reduction precision ($\sim1.6$ m/s) for mid to late K dwarfs. For somewhat bluer stars, they derived slightly higher astrophysical jitter levels ($\sim1.4$ m/s), but still only a weak correlation with chromospheric activity.       

It seems that in the \harps high precision data, the lower boundary of the minimal total rms is of the order of 50 cm/s for stars with parameters mentioned above, and our current reduction pipeline. As gaining insight in the fundamental limiting factors for the radial velocity technique is  an important task, we will present an in-depth analysis of the statistical properties of the sample regarding RV jitter in a dedicated  forthcoming paper (Lovis et al. in prep.).

For the work presented here, we restrict ourselves to a pragmatic approach and compare the three stars with similar ones in the sample for which the jitter can be estimated in form of the rms of the residuals to the fitted orbits, $\sigma(O-C)$ (this number can however also contain additional contributions e.g. from undetected low mass planets). Both  HD\,85390 (K1V) and HD\,103197 (K1V) are similar to \object{HD\,69830} (K0V) in terms of  spectral type, effective temperature and  $\log{R^{'}_{HK}}$, even though that  HD\,85390 is somewhat cooler than this star. As mentioned, for  HD\,69830 a $\sigma(O-C)$ of 0.64 m/s was found in Lovis et al. (\cite{lovisetal2006}) for the later measurements. Note that since then, our calibration algorithm has been improved even more (Pepe \& Lovis \cite{pepelovis2008}). HD\,90156 (G5V) is similar to \object{HD\,47186} (G5V) for which a total $\sigma(O-C)$ of 0.91 m/s was found in Bouchy et al. (\cite{bouchyetal2009}). From this we conclude that the typical levels of the unquantified errors are of order of 0.5 to 0.8 m/s, where the latter value was already used in Pepe et al. (\cite{pepeetal2007}).

Following Saar et al. (\cite{saaretal1998}) or Pepe et al. (\cite{pepeetal2007})  we then assume that we can quadratically add this estimated radial velocity jitter to the quantified error of each radial velocity measurement, using a constant value of either 0.5 or 0.8 m/s. This might not be completely justified as depending on the physical nature of the variability source like magnetic cycles of the star (Santos et al. \cite{santosetal2010}), the jitter might be rather of a red noise type, and varying in time. Using such an additive procedure, we can however certainly mitigate the adverse effects of artificially high dynamics in the quantified (photonic) errors.  We then search for orbital solution in the data after adding the two mentioned values, and compare the impact on the best fit parameters and their errors.

\subsection{HD\,85390: An intermediate mass planet and a long-term drift}

HD\,85390 was monitored with \harps for a time span of 6.5 years (2386 days), during which 58 measurements were obtained. Except for the first six measurements we always followed the observational strategy to use a fixed exposure time of 15 minutes in order to minimize the effects of stellar acoustic modes (Pepe \& Lovis \cite{pepelovis2008}).  We however nevertheless include the early, less accurate measurements to have a better constraint on the long time drift present in the data, cf. next. The spectra which were all obtained in ThAr simultaneous mode have typical S/N at $\lambda=550$ nm of 116  and a mean quantified error of the radial velocity (including photon noise, calibration and instrumental drift uncertainty) of 0.62 m/s.  Given the low chromospheric activity of the star, they should however induce a clearly smaller raw rms than the one found in the 58 measurements which is 3.3 m/s (peak-to-peak 14.4 m/s), indicating instead the presence of a companion. To account for the unquantified errors (see section \ref{sec:jitter}), we add quadratically to all our measurements an additional error of 0.5 resp. 0.8 m/s. Using either 0.5 m/s or 0.8 m/s has however only very minor influences on the orbital solutions: the derived quantities like the period $P$, the mean longitude $\lambda$, the eccentricity $e$ or the velocity semi-amplitude $K$ agree to better than  2 \% for the two cases,  much less than the errors bars on these quantities. This is in good agreement with Wright (\cite{wright2005}) who states that radial velocity jitter is more important in the detection of extrasolar planet than in their characterization, and that the effects of including jitter in the noise estimation for the derived orbital parameters of the best fit is typically  small, at least for a case like here where the induced radial velocity signal is larger than the estimated jitter amplitude.

\begin{figure}[th!]
   \includegraphics[angle=0,width=0.5\textwidth,origin=br]{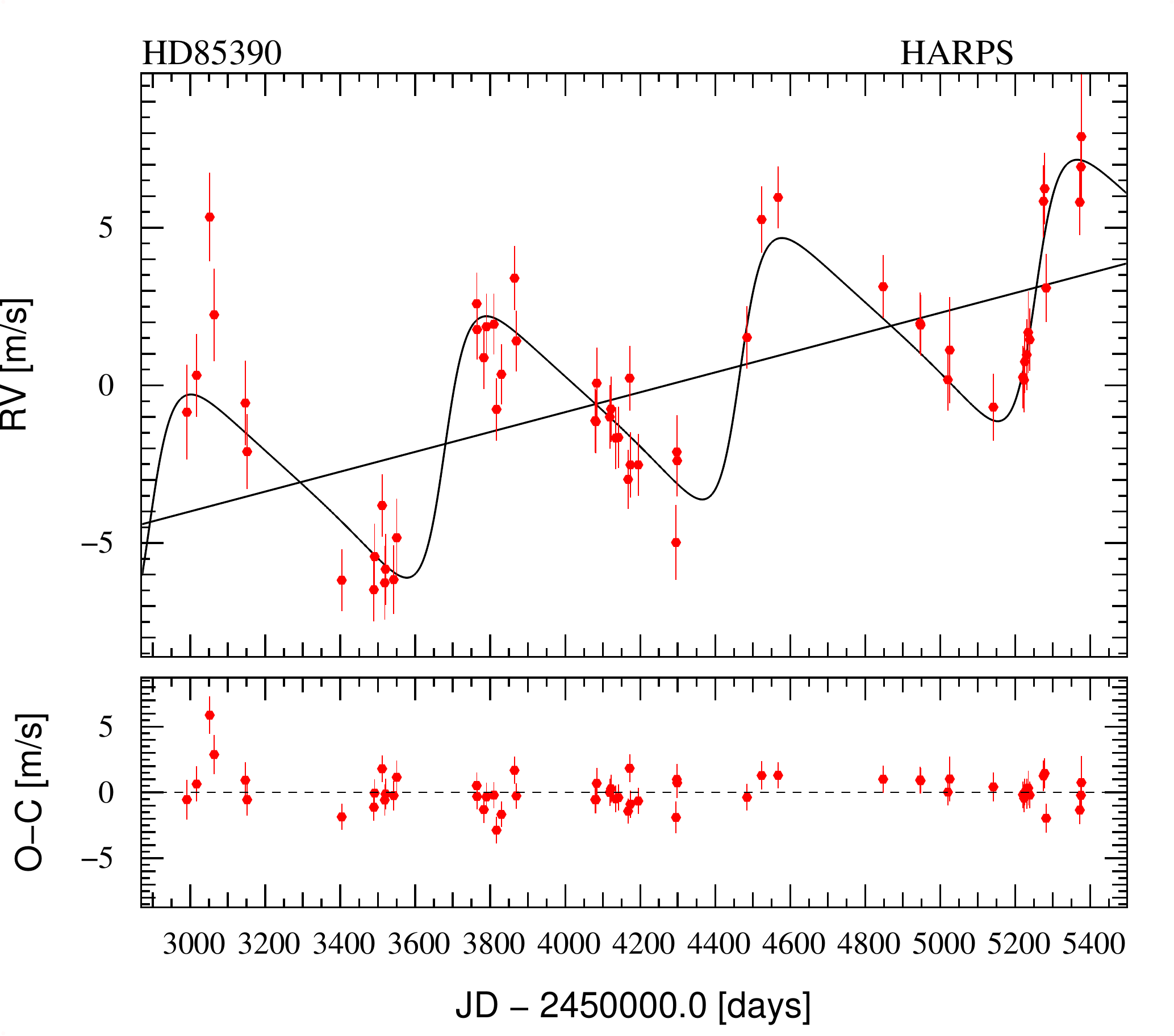}
   \includegraphics[angle=0,width=0.5\textwidth,origin=br]{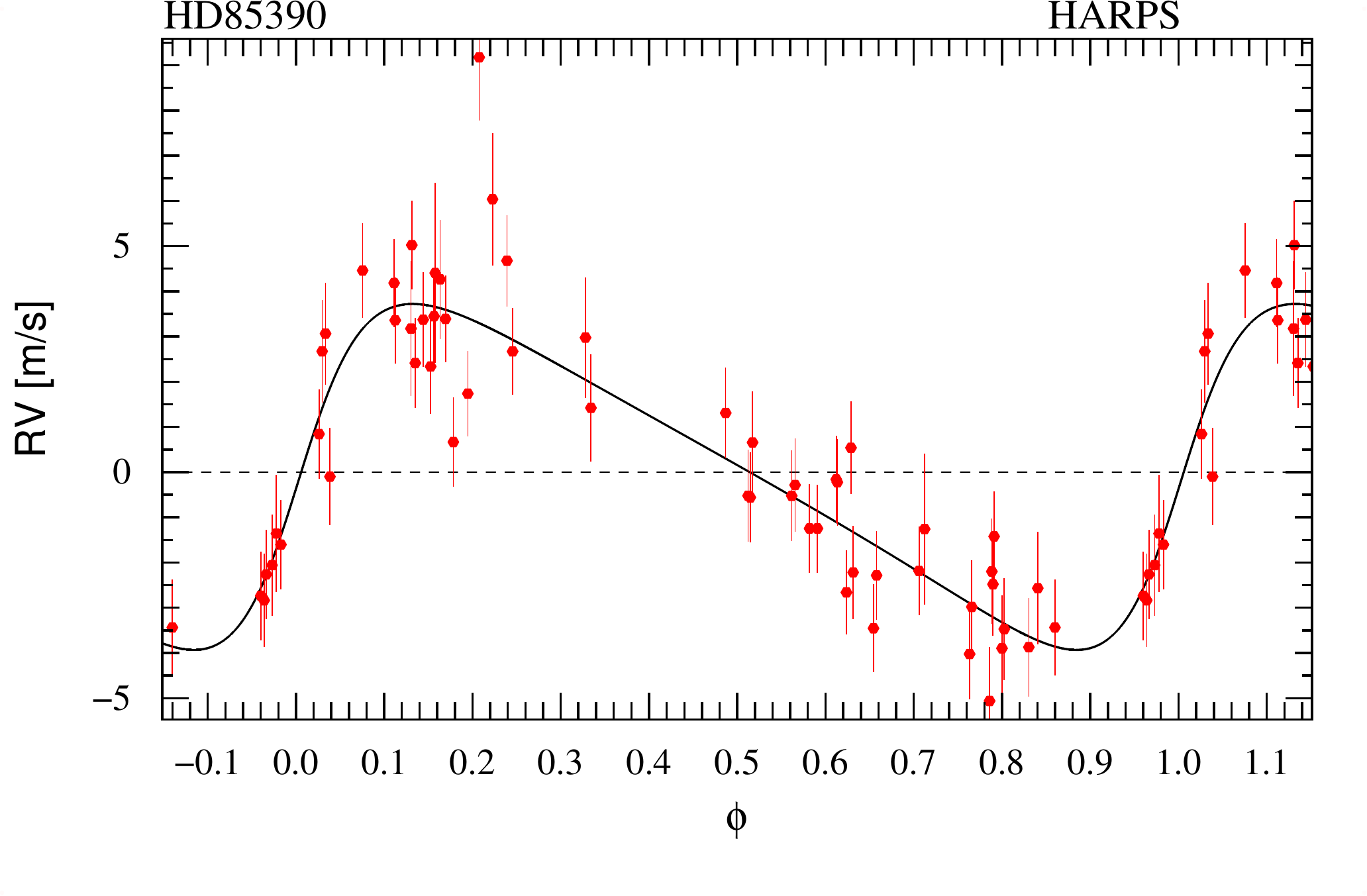}
   \includegraphics[angle=0,width=0.5\textwidth,origin=br]{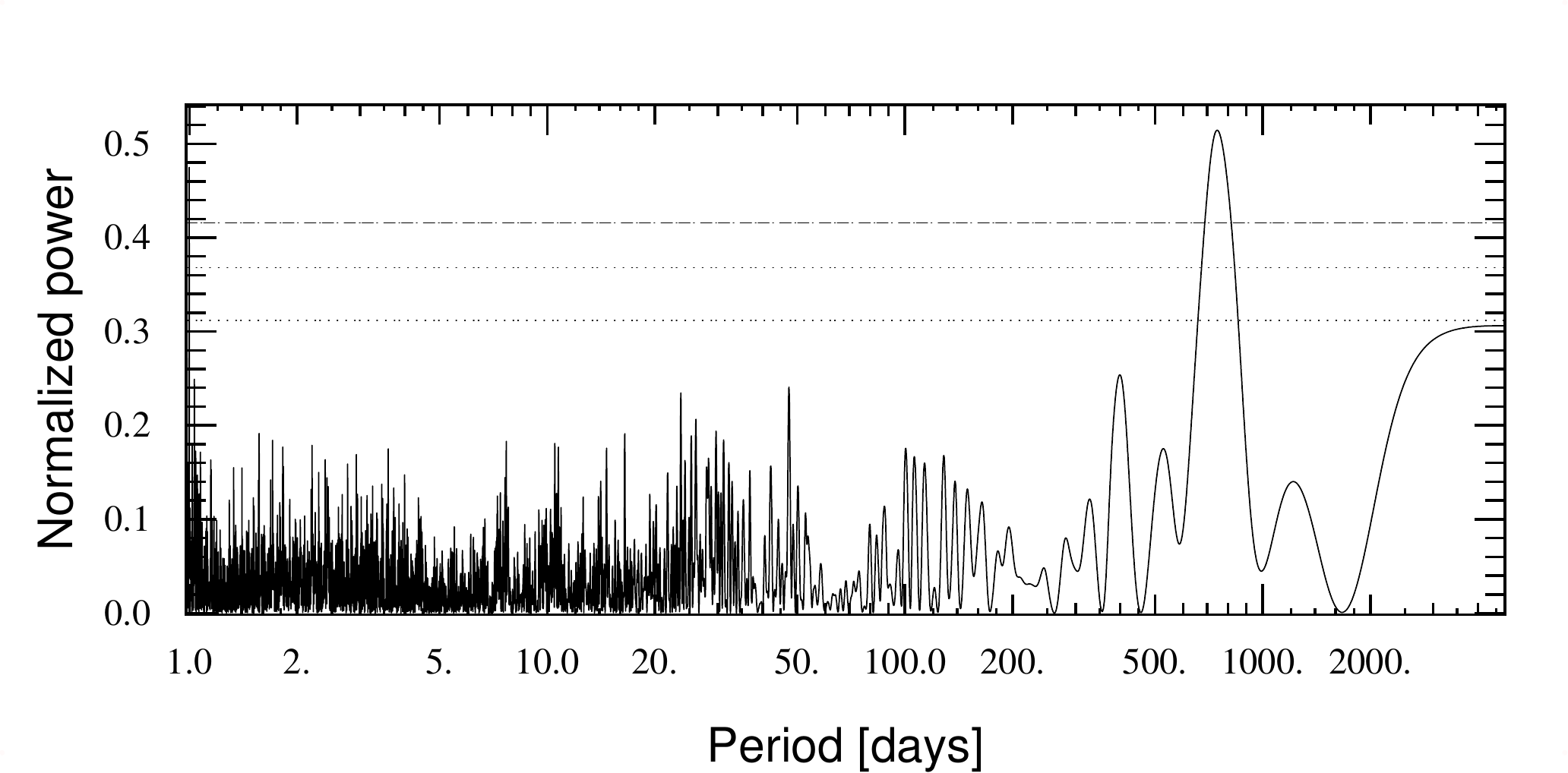}
 \caption{Radial velocity measurements of HD\,85390 as a function of Julian Date (top). The six earliest measurements have larger error bars. The best solution with a linear drift plus one Keplerian orbit is also displayed as well as the residuals to this model. The middle panel shows the radial velocity as a function of the orbital phase for HD\,85390\,b, with the effect of the drift removed. The bottom panel shows the generalized Lomb Scargle periodogram of the observations. The peak at a period of about 800 days is very prominent, and has a false alarm probability (FAP) of much  less than 0.1 \%. The three horizontal lines indicate FAPs of 0.1, 1 and 10\% from top to bottom.}\label{fig:HD85390}
\end{figure}

Radial velocity measurements as a function of Julian date (with 0.8 m/s of additional error) are shown as red dots in the top panel of Fig. \ref{fig:HD85390}. A long-term, linear drift overlaid by a higher frequency modulation with a period of about 800 d can be seen directly in the data. Of the latter, three phases are covered.  The bottom panel of Fig. \ref{fig:HD85390}  shows the generalized Lomb Scargle periodogram (as described in Zechmeister \&  K\"urster \cite{zechmeisterkurster2009}) of the data. The peak at such a period is clear and has a very low FAP of  less than 0.01\%, as found by scrambling the data 10\,000 times which confirms the visual impression of a very clearly detected orbit. The FAPs are obtained by performing random permutations of the velocities, calculating the periodogram, recording the peak power for each trial, and finally comparing the power of the real signal to the peak power distribution of the permuted datasets.

Therefore we fit the measurements with a linear drift and a Keplerian orbit, using a genetic algorithm (see below for a discussion of the nature of  the drift). The resulting best fit and the residuals around it are also plotted in the top panel of Fig. \ref{fig:HD85390}. The phase-folded radial velocity curve of the planet candidate after removing the drift is shown in the middle panel of this figure. The reduced $\chi_{r}^{2}$ per degree of freedom is 1.52 (2.31 for an additional error of 0.5 m/s) and the weighted r.m.s. of the residuals around the solution $\sigma(O-C)$ is 1.15 m/s (1.14 m/s for an additional error of 0.5 m/s), which is not much more than the mean total assumed error of the individual measurements.  In the residuals, no additional significant periodic signals is found.  

The somewhat higher  $\sigma(O-C)$ is also at least partially due to the mentioned six early measurements which can be identified in Fig. \ref{fig:HD85390} by their larger error bars. Excluding these six early radial velocities results in a very similar fit with a reduced $\chi_{r}^{2}=1.10$ (additional error of 80 cm/s) and reduces  $\sigma(O-C)$ to just 1.00 m/s.

The long term signal is found to have slope of 1.15$\pm$0.02 m/s yr$^{-1}$. The derived  parameters for the Keplerian orbit imply a period $P=788\pm25$ d (corresponding to a semi-major axis $a=1.52$ AU), a minimum mass of $\msini=42.0\pm3.6\, \mearth$ ($2.44\,\mnept$), and eccentricity $e=0.41\pm0.12$. The error intervals were computed for a  68\% confidence level with 5000 Montecarlo iterations, where the actual measurements are replaced with a value drawn from a normal distribution around the nominal value with a standard deviation which is the same as the error bar on the point. These altered data sets are then again fitted to obtain the confidence intervals from the distribution of orbital elements one finds.

The orbital elements are listed in Table \ref{tab:HD85390solution}. We checked that there is no correlation between bisector shape of the cross-correlation function and the radial velocity or the residuals of the Keplerian fit,  providing strong support to the planetary interpretation of the RV signal.

\begin{table}
\caption{Orbital and physical parameters of the 1-planet plus linear drift solution of the {\footnotesize HARPS} data for  the objects around  \object{HD\,85390}.  $\lambda$ is the mean longitude at the  barycenter of the observations at BJD 54269.3584. }\label{tab:HD85390solution}
\begin{center}
\begin{tabular}{llccc}
\hline\hline
Parameters          &                                      &      \object{HD\,85390\,b}       \\\hline                                                                                                                  
$P$                       &  [days]                         &      $788 \pm 25$                \\ 
$\lambda$                 & [deg]         &      $177\pm25$              \\      
$e$                       &                                      &     $0.41\pm0.12$      \\  
 $\omega$           & [deg]                            &     $-94 \pm23$          \\  
$K$                      & [m/s]                            &   $3.82\pm0.33$        \\  
$V$                         &  [km/s]                         & $33.0853\pm0.0004     $    \\                                                             
slope                      &[m/s yr$^{-1}$]                &   $1.15\pm0.02 $                      \\             
$f(m)$              & [$10^{-9}\, M_{\odot}$  ]   & $0.0035\pm0.0009 $                       \\
$\msini$  & [$\mearth$]                    & $42.0 \pm 3.6 $                \\
$a$                      & [AU]                                  & $1.52\pm 0.04 $             \\\hline                                                                
$N_{\rm meas}$            &                                      & 58                          \\
span             & [years]                          &  6.5                          \\
$\sigma(O-C)$    & [m/s]                         & 1.15                     \\         
$\chi_{r}^{2}$            &                            & 1.52                     \\   
\hline
\end{tabular}
\end{center}
\end{table}

\begin{figure}[t]
   \includegraphics[angle=0,width=0.5\textwidth,origin=br]{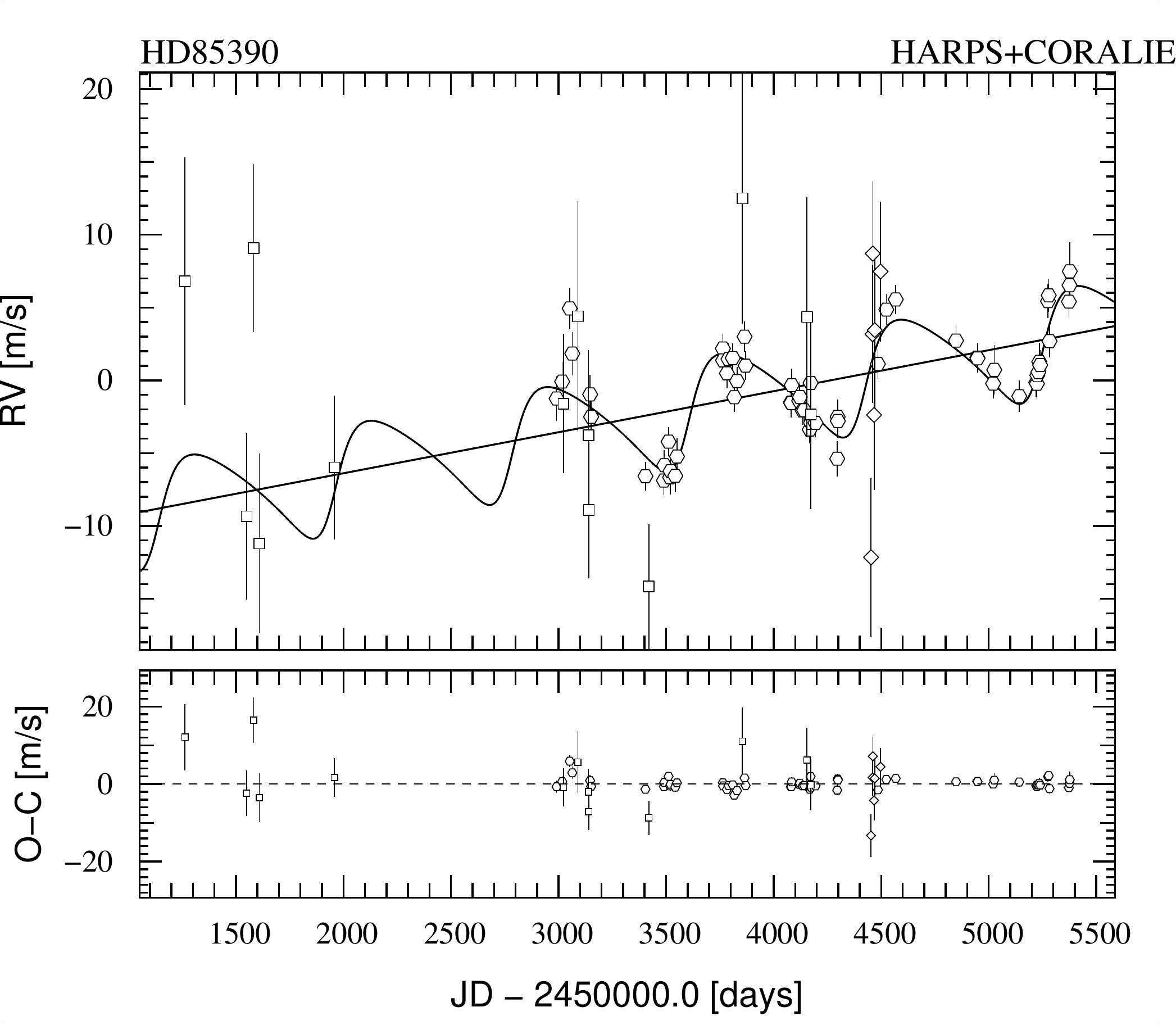}
   \caption{Radial velocity measurements of HD\,85390 as a function of Julian Date for the combined data set of {\footnotesize CORALIE} (squares for measurement before its upgrade in June 2007, and diamonds after this moment) and \harps (circles), together with a one Keplerian plus linear drift fit (solid line). The small panel shows the residuals around the fit. }\label{fig:HD85390harpscoralie}
\end{figure}

HD\,85390 is also part of the  {\footnotesize CORALIE} planet search survey (Udry et al. \cite{udryetal2000}). Including the 19 measurements made with {\footnotesize CORALIE} expands the observed time base significantly to 11.3 years (4114 days), back to JD=2\,451\,262 which could help to better constrain the long-term drift seen in the \harps data. The radial velocity as a function of time for the two instruments is shown in Fig. \ref{fig:HD85390harpscoralie}. It is found that at the level of precision of {\footnotesize CORALIE} (about 6.5 m/s for this target), the star is stable, i.e. the ratio of the rms of all the velocity measurement and the mean error of one velocity measurement is about unity. Repeating the fitting process with the combined data set leads to a solution for the Keplerian orbit which agrees within the error bars  with the solution without the  {\footnotesize CORALIE} data,  namely $P=821\pm20$ d, $e=0.34\pm0.10$ and $\msini=42.3\pm3.5\,\mearth$, with a $\chi_{r}^{2}$ of 1.59 for additional jitter of 0.8 m/s in the {\footnotesize HARPS} measurement. The linear component of the drift is now $1.03\pm0.02$ m/s yr$^{-1}$.  From that we can exclude the presence of a massive outer companion with a period smaller than about 10 years.  

In the nominal model for the orbital solution, we have assumed a Keplerian orbit plus a linear drift. One wonders if we could significantly improve the fit by allowing for a quadratic component in the drift, as one could tentatively infer from two early  {\footnotesize CORALIE} measurement lying 12 resp. 16 m/s above the linear fit.

This could be indicating the presence of a body that might be interesting to direct imaging searches. To explore this significance, we have fitted the combined dataset of \harps and {\footnotesize CORALIE} which is more relevant in this context as it constrains better the long time behaviour also with a Keplerian plus a quadratic drift. The orbital solution for the Keplerian orbit corresponds then to a companion  with properties that again agree within the error bars with the solution shown in table \ref{tab:HD85390solution}. This is due to the fact that the \harps measurements alone constrain this orbit. The reduced $\chi_{r}^{2}$ for the combined data are now 1.47 for an additional errors of 0.8 m/s, i.e. somewhat less than for the linear model with one degree of freedom less. We have then used an F-test as in Pourbaix \& Arenou (\cite{pourbaixarenou2001}) to get the statistical probability that the improvement due to the quadratic model is significant.  We find a probability of  98.7 \%, corresponding to about 2.5 sigma. The main assumption behind the F-Test is however that the model is linear and that the error distribution is Gaussian, which is both not strictly fulfilled, therefore this number must be regarded with caution. Thus, and because the orbital solution of HD\,85390\,b itself is not significantly affected by the choice, we prefer to stick for the moment to the simpler linear model and  conclude that to constrain further  the nature of the long term drift, more high precision observations with \harps are needed.

Long term, small amplitude signals could in principle also be due to magnetic cycles of the host star. While we cannot absolutely exclude this possibility, we think that it is unlikely in the present case, due to the following reasons: First,  during the observed time span, $\log{R^{'}_{HK}}$ decreases linearly from about -4.93 to -5.00. Thus, the change of $\log{R^{'}_{HK}}$ and the long term drift in the radial velocity are clearly anti-correlated. This is the contrary of what is expected, as in active zones, convection tends to be inhibited, reducing the convective blue shift, so that a shift of the radial velocity into the red should occur (Santos et al. \cite{santosetal2010}). Second, the change of  $\log{R^{'}_{HK}}$, about 0.07 dex is less than the value of about 0.1 dex where we start to see clear correlations of the activity level and the radial velocity in the \harps database (Lovis et al., in preparation). Third, the amplitude of the drift is large ($\sim8$ m/s), much more than expected for the small change in  $\log{R^{'}_{HK}}$ for a K dwarf.

\subsection{HD\,90156: A warm Neptune analogue}
Since the mentioned change of strategy to minimize stellar noise (fixed exposure times of 15 minutes), we have gathered 66 radial velocities in ThAr simultaneous mode during a time period of  1614 days (4.4 years). HD\,90156 is a bright star (V=6.92), therefore the measurements have a  high mean S/N of  246 at 550 nm, with variations of the S/N between 121 and 342, depending mostly on seeing conditions. The derived mean quantified error (photon noise, calibration and drift uncertainty) is 0.41 m/s.  The observed raw rms for the 66 measurements is 2.6 m/s, again more than expected for the low stellar activity level. As before, we add quadratically to all measurements an additional error of either 0.5 or 0.8 m/s and check for the consequences. 

\begin{figure}
   \includegraphics[angle=0,width=0.5\textwidth,origin=br]{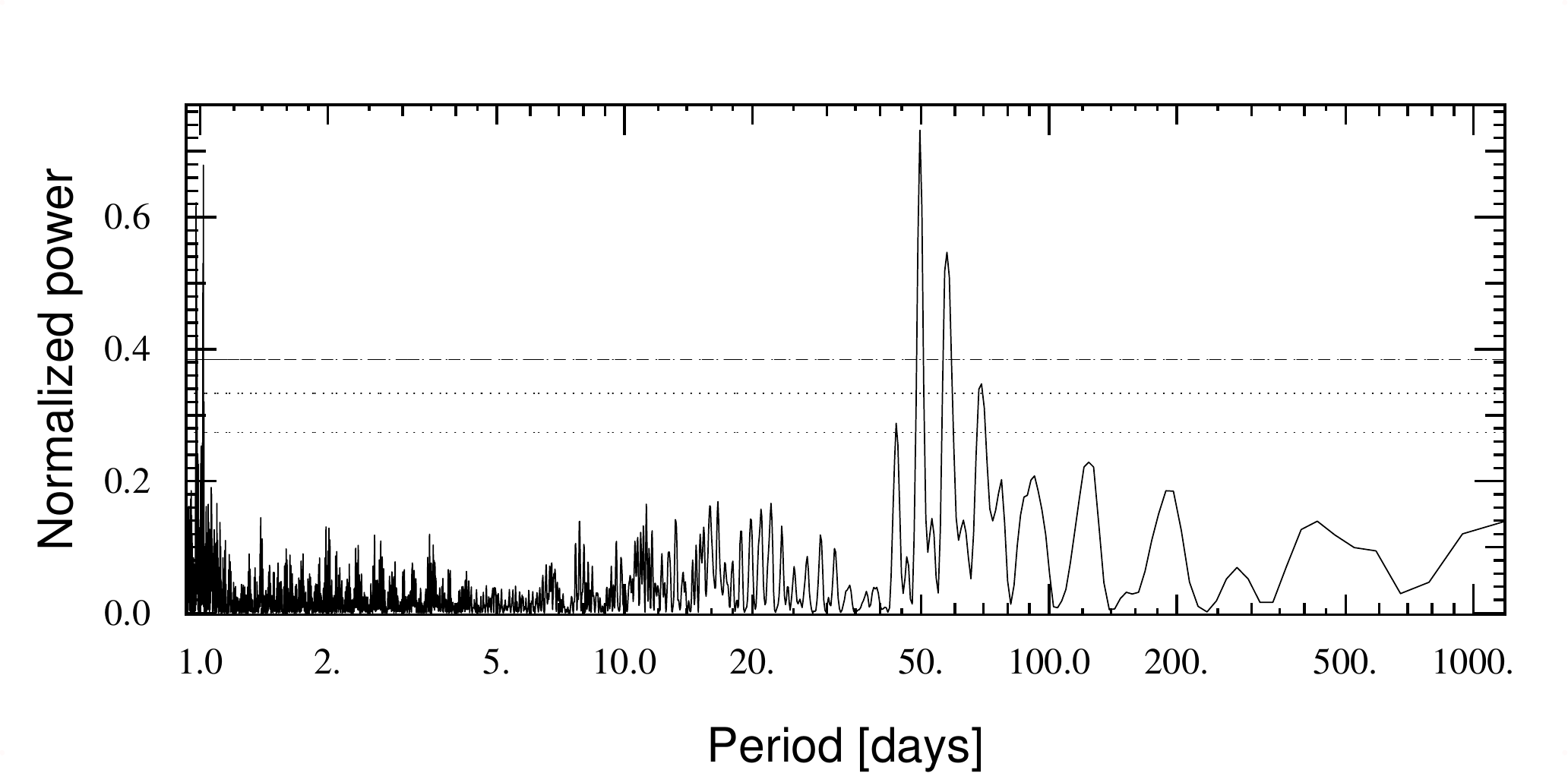}
 \includegraphics[angle=0,width=0.5\textwidth,origin=br]{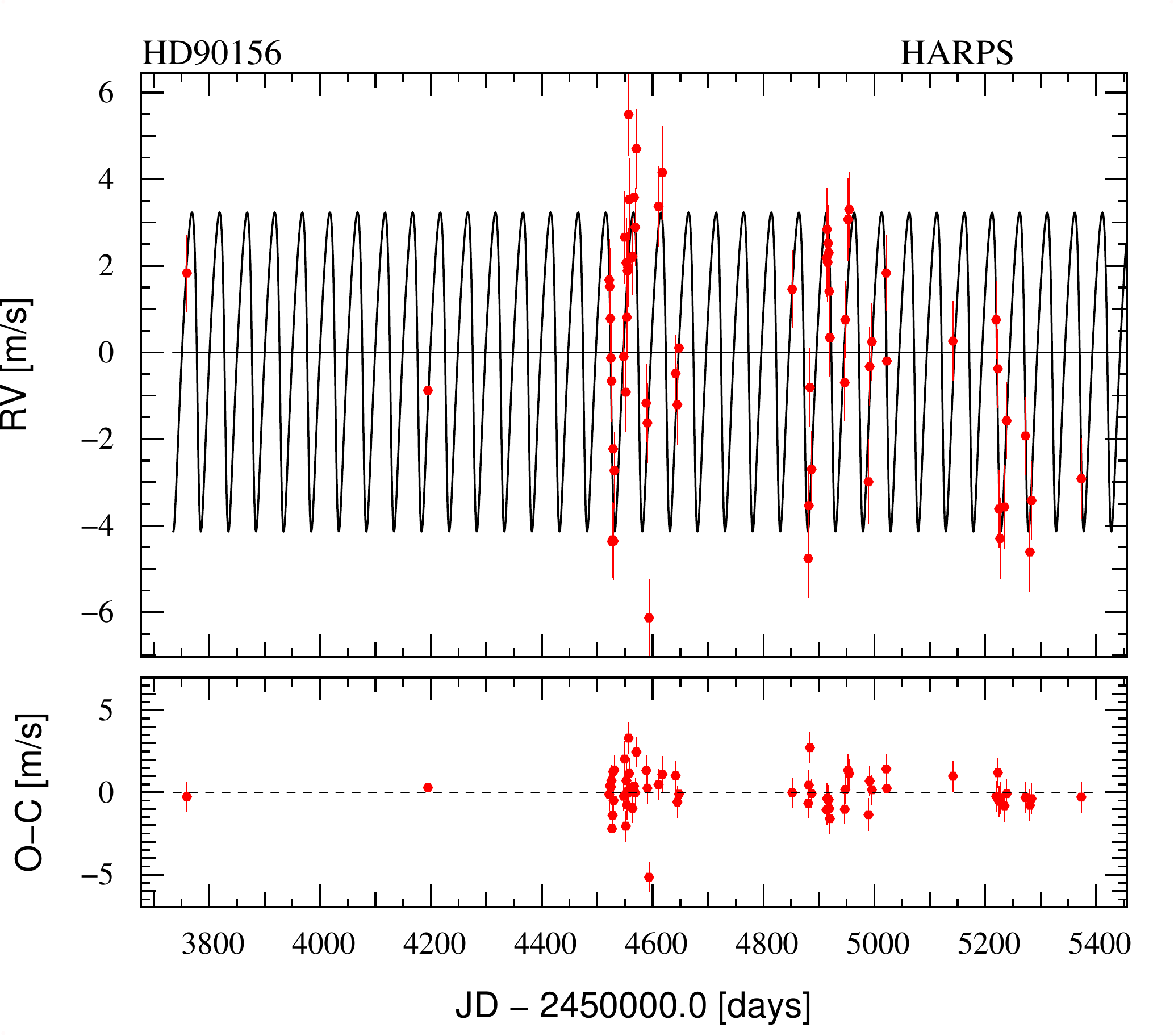}
   \includegraphics[angle=0,width=0.5\textwidth,origin=br]{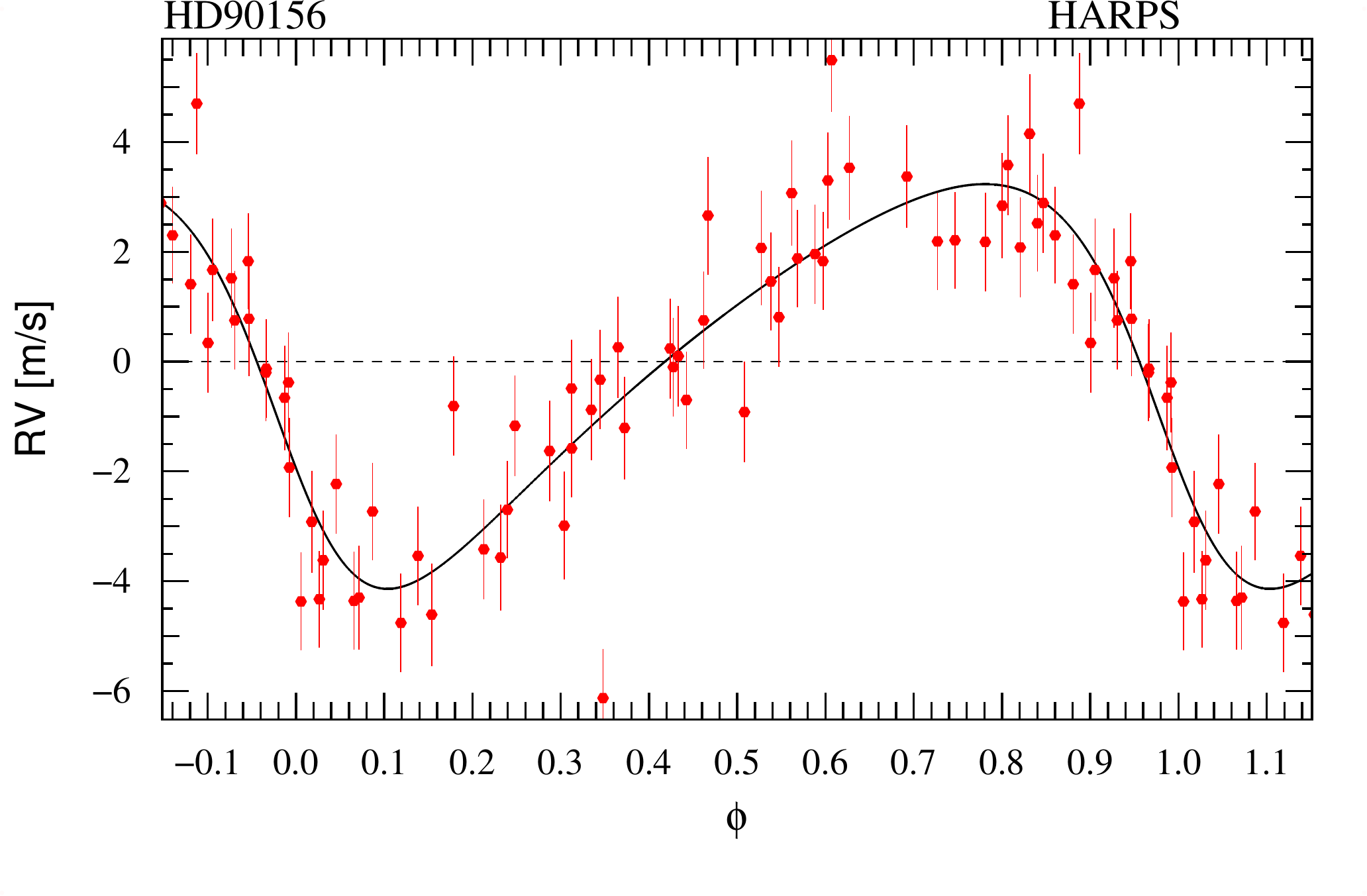} 
\caption{Top panel: Generalized Lomb Scargle periodogram of the radial velocity measurements of HD\,90156. A FAP of 0.1\% is indicated by the dashed-dotted line. The peak at about 50 days is very clear and has a FAP  much less than this value.  Middle panel: Radial velocities and residuals as a function of Julian Date, and (bottom panel) as a function of orbital phase for the one-planet Keplerian fit ($P=49.8$ days).}\label{fig:HD90156}
\end{figure}

A periodogram of the data which is displayed in the top panel of Fig. \ref{fig:HD90156} shows a very strong peak at a period of about 50 days with a FAP of much less than 0.1\% as obtained by executing 10\ 000 data scrambling experiments. A number of aliases can also been seen. A one Keplerian fit assuming an additional RV error of 0.8 m/s quickly converges on a solution with a $\msini=17.98\pm1.46\,\mearth$ planet on an orbit with a period of $49.77\pm0.07$ days and an eccentricity of $0.31\pm 0.10$. This solution and the measurements are shown in  the lower two panels of Fig. \ref{fig:HD90156}.  The reduced  $\chi_{r}^{2}$  is 2.03, and the  $\sigma(O-C)$ is 1.23 m/s (see Table \ref{tab:HD90156}). The mentioned period corresponds to a semi-major axis of  0.25 AU, and the projected mass is very close to the mass of Neptune ($1.05\,\mnept$), making this planet a warm Neptune analogue. We checked that there is no correlation between the bisector shape of the cross-correlation and the radial velocity or the residuals around the Keplerian fit. Assuming an additional error of 0.5 instead of 0.8 m/s leads to virtually identical fit parameters with relative differences of less than 3 \%. The reduced  $\chi_{r}^{2}$  increases to 3.91, indicating that we now underestimate the unquantified errors in the measurements. We note the subsolar metallicity of the parent star, so that \object{HD\,90156\,b} is another example of a Neptunian planet not following the ``metallicity effect'' seen for giant planets, as discussed for example in  Sousa et al. (\cite{sousaetal2008}). It is also an example of a warm low mass planet which does not show the same strong pile-up behaviour at about 0.04 AU as higher mass Hot Jupiters do (Udry \& Santos \cite{udrysantos2007}).

\begin{table}
\caption{Orbital and physical parameter of the 1-planet Keplerian solution for the planet around  \object{HD\,90156}. The  mean longitude $\lambda$ is given at the barycenter of the observations, at JDB 54772.0784.}\label{tab:HD90156}
\begin{center}
\begin{tabular}{llccc}
\hline\hline
Parameters          &                                      &      \object{HD\,90156\,b}       \\\hline                                                                                                                  
$P$                       &  [days]                         &      $49.77 \pm 0.07$                \\ 
$\lambda$                 &[deg]         &      $91\pm15 $             \\      
$e$                       &                                      &     $0.31\pm0.10 $     \\  
 $\omega$           & [deg]                            &     $113 \pm18 $          \\  
$K$                      & [m/s]                            &   $3.69\pm0.25 $        \\  
$V$                         &  [km/s]                         & $27.0273\pm0.0003 $        \\                                                             
$f(m)$              & [$10^{-9}\, M_{\odot}$  ]   & $0.00022\pm0.00005 $                       \\
$\msini$  & [$\mearth$]                    & $17.98 \pm 1.46 $               \\
$a$                      & [AU]                                  & $0.250\pm 0.004  $            \\\hline                                                                
$N_{\rm meas}$            &                                      &  66                         \\
span             & [years]                          &     4.4                          \\
$\sigma(O-C)$    & [m/s]                         &  1.23                     \\         
$\chi_{r}^{2}$            &                            & 2.03                     \\   
\hline
\end{tabular}
\end{center}
\end{table}

In a periodogram of the residuals around the one planet fit, there is a signal at a period of about 14 days. With a FAP of 16 \%  we can however not yet conclude anything about its origin. We however hope to soon be able to either confirm or reject this signal by additional observations.

\subsection{HD\,103197: An intermediate mass planet}
\begin{figure}
 \includegraphics[angle=0,width=0.5\textwidth,origin=br]{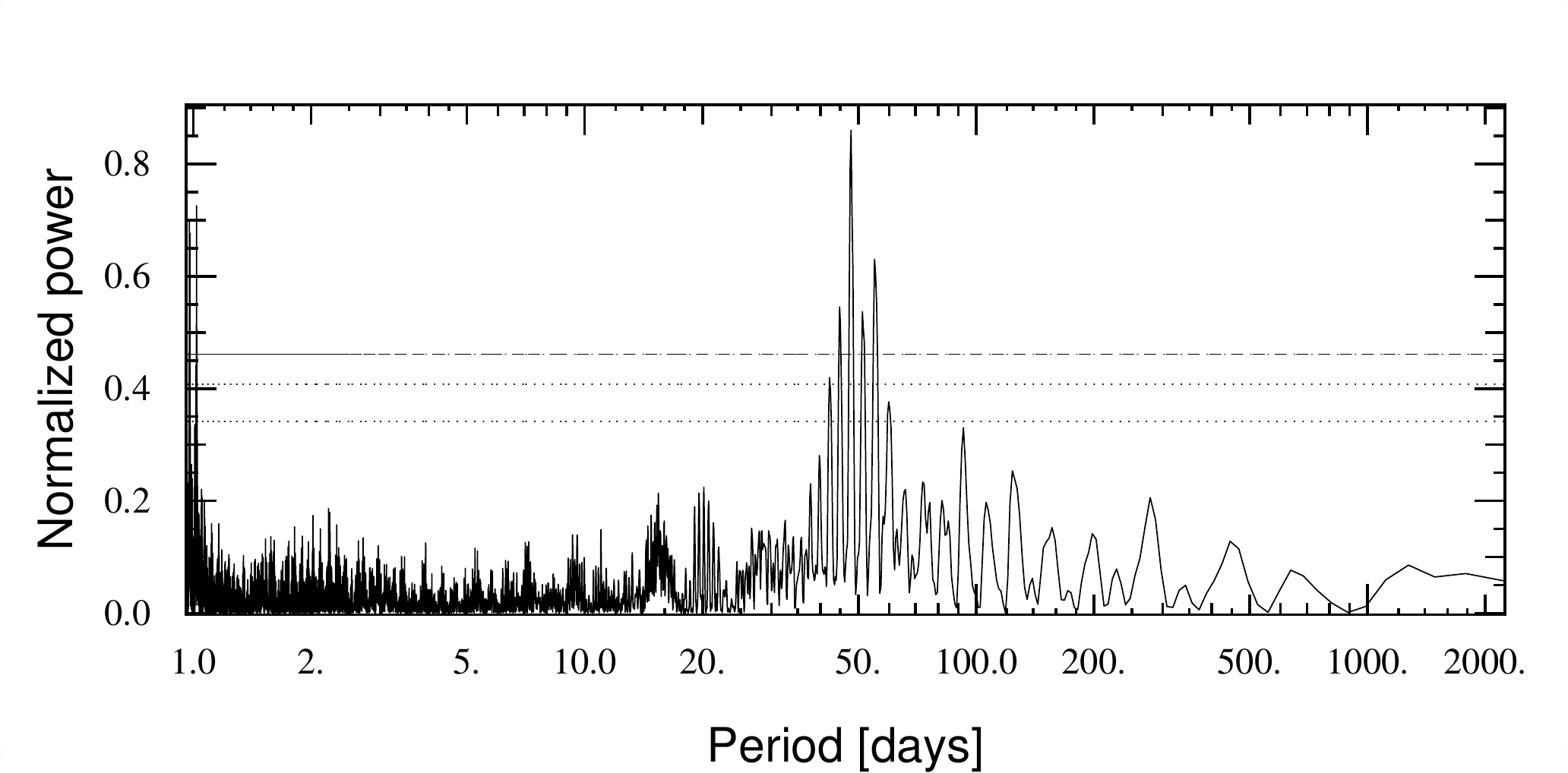}
   \includegraphics[angle=0,width=0.5\textwidth,origin=br]{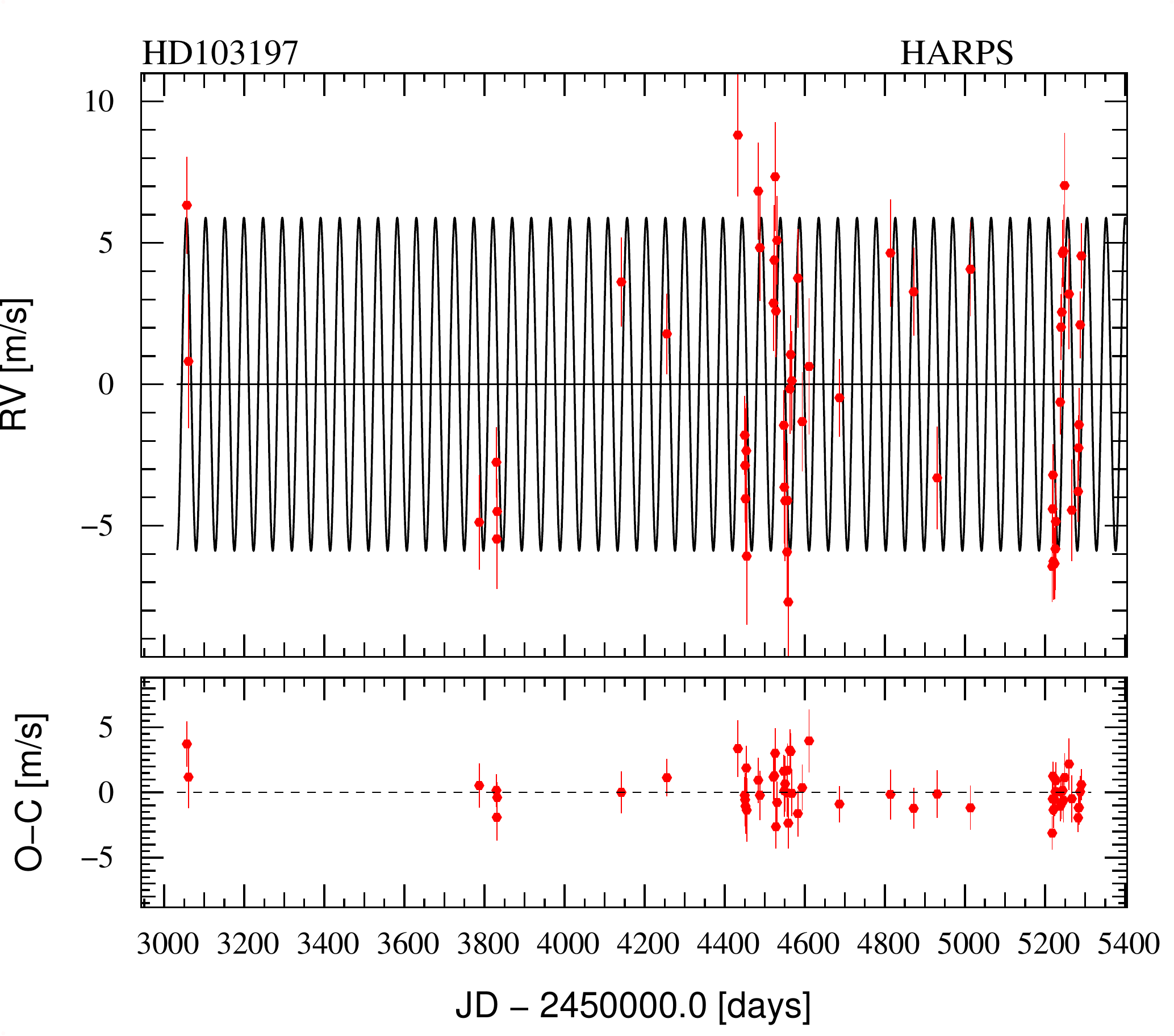} 
      \includegraphics[angle=0,width=0.5\textwidth,origin=br]{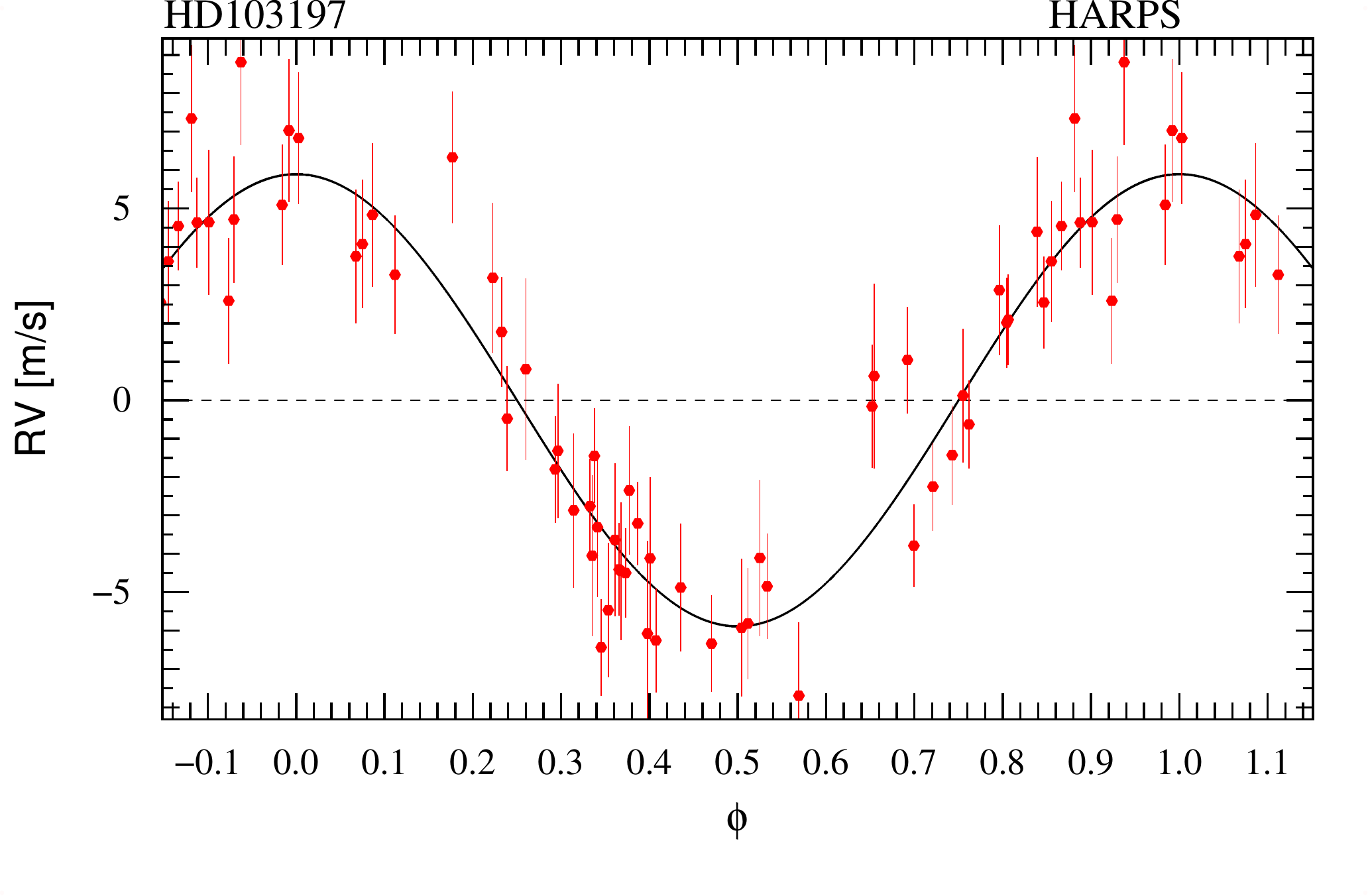} 
  \caption{Top panel: Generalized Lomb Scargle periodogram of the 58 radial velocity measurements of HD\,103197. The middle and bottom panel show the radial velocity and the residuals as a function of Julian Date, and as a function of orbital phase together with  the orbital solution  given in table \ref{tab:HD103197}.}\label{fig:HD103197}
\end{figure}

We have gathered 58 high quality radial velocity measurement of HD\,103197 over a time period of 6.12 years. They are shown in the top panel of Fig. \ref{fig:HD103197}. The mean exposure time for this star was shorter than for the other three, as it does not belong to the high precision sample, and of the order of 5 minutes, which should however still be just sufficient to mean out the most important stellar oscillation period for this type of star (cf. Pepe \& Lovis, \cite{pepelovis2008}). Corresponding to the smaller exposure time, the spectra have lower typical S/N at 550 nm, namely between 33 and 107 with a mean of 59. The error on a single radial velocity measurement which includes photon noise, calibration and spectrograph drift uncertainty varies between 0.7 and 2.3 m/s, with a mean of 1.1 m/s. The observed raw rms of the radial velocities is with 3.95 m/s clearly larger, and indicates the presence of a companion. As before, we add quadratically additional errors of 0.8 and 0.5 m/s. For this star, no ThAr simultaneous reference mode was usually used, which is however unproblematic given the very small nightly drift of the instrument and the fact that we mix observations with and without  ThAr  during one night. 

A periodogram of our radial velocity data shows a very strong peak at a period of about 48 days, as can be seen in the top panel of Fig. \ref{fig:HD103197}, where horizontal lines again indicate FAPs of $10^{-1}$, $10^{-2}$ and $10^{-3}$ from bottom to top. Fitting the data with one Keplerian orbit leads to a solution with an insignificant eccentricity ($0.04\pm0.05$) so that we fix it to zero, leaving us with four free parameters.

The following solution is then found for the companion, assuming a additional radial velocity jitter of 0.8 m/s: An orbital period of $47.84\pm0.03$ days, a mean longitude $\lambda=151\pm3\deg$ at JDB 54846.0994 and $K=5.9\pm0.3$ m/s,  implying a minimum mass $\msini= 31.2\pm2.0\,\mearth$  and a semi-major axis $a =0.25$ AU.  The mass corresponds to 1.8 $\mnept$. The fit has a reduced $\chi_{r}^{2}$ of 1.02 and  $\sigma(O-C)=1.40$ m/s, which corresponds well with the total assumed mean error of the individual measurements. As for the two previously discussed stars, using an additional jitter of 0.5 m/s instead of 0.8 m/s has only very minor influences on the fit, and results in a $\chi_{r}^{2}=1.18$. The middle and lower panel of Fig. \ref{fig:HD103197} show the radial velocities and the fit as a function of time, and phase folded, as well as the residuals around the fit. The orbital elements for the Keplerian fit with $e=0$ are given in Table \ref{tab:HD103197}. The errors were again derived from 5000 Monte Carlo simulations.

\begin{table}
\caption{Orbital and physical parameter of the  1-planet Keplerian solution for the companion around  \object{HD\,103197}. The eccentricity of the planet is fixed to zero. The epoch of $\lambda$ is JDB 54846.0994.}\label{tab:HD103197}
\begin{center}
\begin{tabular}{llccc}
\hline\hline
Parameters          &                                      &  \object{HD\,103197\,b}      \\\hline                                                                                                                  
$P$                       &  [days]                         &  $47.84\pm0.03$               \\ 
$\lambda$                 &[deg]         &   $151\pm3$                 \\      
$e$                       &                                      &  0.0 (fixed)     \\  
 $\omega$           & [deg]                            &  0.0 (fixed)          \\  
$K$                      & [m/s]                            &  $5.9\pm0.3$         \\  
$V$                         &  [km/s]                         & $-4.3282\pm0.0002$         \\                                                             
$f(m)$              & [$10^{-9}\, M_{\odot}$ ]   & $0.0010\pm  0.0002$                        \\
$\msini$  & [$\mearth$]                    &$31.2\pm2.0$                   \\
$a$                      & [AU]                                  &   $0.249\pm0.004$               \\\hline                                                                
$N_{\rm meas}$            &                                      &    58                      \\
span             & [years]                          &     6.12                        \\
$\sigma(O-C)$    & [m/s]                         &  1.4                   \\         
$\chi_{r}^{2}$            &                            &  1.02                     \\   
\hline
\end{tabular}
\end{center}
\end{table}

We note that the orbital period is similar to the rotation period of the star $P_{\rm rot}$ which we estimate to be $51\pm5$ days. We however checked for periodicity in the bisector shape at such a period, and for a significant correlation of the velocity or the residual and the bisector shape, finding no such signs. We finally looked at the periodogram of the residual around the fit and discovered no additional significant signal.

\section{Summary}\label{sec:summary}   
We have reported in this paper the discovery of three extrasolar planet candidates discovered with the  {\footnotesize HARPS}  Echelle spectrograph mounted on the 3.6-m ESO telescope located at La Silla Observatory.  \object{HD\,85390\,b} is an intermediate mass planet ($\msini=42.0\,\mearth$) on an orbit with a semi-major axis of 1.5 AU. A drift indicates the presence of an additional outer companion, but we cannot currently determine its nature with our data. \object{HD\,90156\,b} is a warm Neptune analogue ($\msini=1.05\, \mnept$, $a=0.25$ AU).  \object{HD\,103197\,b} is  another intermediate mass planet ($\msini=31.2\, \mearth$). Its semi-major axis is 0.25 AU.  

\section{Discussion: Formation of intermediate mass planets}\label{sec:discussion}   
With intermediate mass planets we here refer to planets with a mass larger than Neptune, but less than Saturn. They therefore fall in the significant mass gap of a factor 5 between Uranus ($17.2\,\mearth$) and Saturn ($95.2\,\mearth$) which corresponds to an important difference in internal composition, too: Uranus and Neptune are ice giants which consist mainly of a heavy elements (iron, rocks and ices) and only about 5-15\% in mass of hydrogen and helium (e.g. Figueira et al. \cite{figueiraetal2009}). Saturn and Jupiter are in contrast gas giants with only a small fraction of heavy elements. Instead, their massive gas envelopes account for most of the mass, namely for about 67 to 80\% for Saturn, and 87 to 97\% for Jupiter (Guillot \cite{guillot1999}).

This difference in mass and composition is in turn understood in theoretical formation models based on the core accretion paradigm as the consequence of an important difference in the formation history of these two groups of planets: Jupiter and Saturn are thought to have undergone a phase of rapid ``runaway'' gas accretion, during which they accreted their large gaseous envelopes on a relatively short timescale, whereas the ice giants never underwent this phase (Pollack et al. \cite{pollacketal1996}).  

If the transformation phase from a Neptunian mass to a Jovian mass is very short  compared to the lifetime of the protoplanetary disk, i.e. if the gas accretion rate onto the planet is very high,  then it is very unlikely that the protoplanetary disk disappears just during the phase, which would obviously terminate the gas accretion and leave the planet at some intermediate mass. We thus see that in this scenario, the absence of a planet with a mass between Uranus and Saturn is not a surprise, but rather an expected outcome (Ida \& Lin \cite{idalin2004}).    

One can therefore say that the higher (smaller) the gas accretion rate in runaway, the smaller (higher) the expected relative frequency of intermediate mass planets compared to giant planets. This correlation will manifest itself in the observed detection rate of intermediate mass extrasolar planets. Indeed point detections made over the last years in particular with \harps to a minimum of the observed planetary mass function at about 30-40 $\mearth$ (e.g. Bouchy et al. \cite{bouchyetal2009}).  Such a minimum (if further confirmed) would be a very important observational finding, as it cannot be explained by an observational bias. This is because the lower mass Neptunian and Super-Earth planets are more difficult to detect than intermediate mass planets. It is interesting to note that a relative paucity of intermediate mass planets seems to exist also in transit searches (Hartman et al. \cite{hartmanetal2009}).

Several physical effects influence the runaway gas accretion rate (Lubow et al. \cite{lubowetal1999}; Ida \& Lin \cite{idalin2004}; Mordasini et al. \cite{mordasinietal2009a}; Lissauer et al. \cite{lissaueretal2009}): Thermal pressure in the planetary envelope at lower planetary masses, the global evolution (dissipation) of the protoplanetary disk which reduces the amount of matter available for the planet, the rate of mass transport due to viscosity within the disk  towards the planet as well as local phenomena like the formation of a gap.  

While several planetary population synthesis simulations built on the core accretion paradigm share the common feature that they predict some depletion of intermediate mass planets relative to other planet types, they attribute in their underlying formation models (cf. Alibert et al. \cite{alibertetal2005}) distinct importance to the various effects mentioned above, so that there are clear differences in the  degree of depletion: from an almost complete absence of such planets between 0.1-1 AU (Miguel \& Brunini \cite{miguelbrunini2008,miguelbrunini2009}), over a significant depletion (Ida \& Lin \cite{idalin2004,idalin2008a,idalin2008b}) to a rather moderate one (about a factor 2-3 relative to Jovian planets) in  Mordasini et al. (\cite{mordasinietal2009a,mordasinietal2009b}).

\begin{figure}
 \includegraphics[angle=0,width=0.43\textwidth,origin=br]{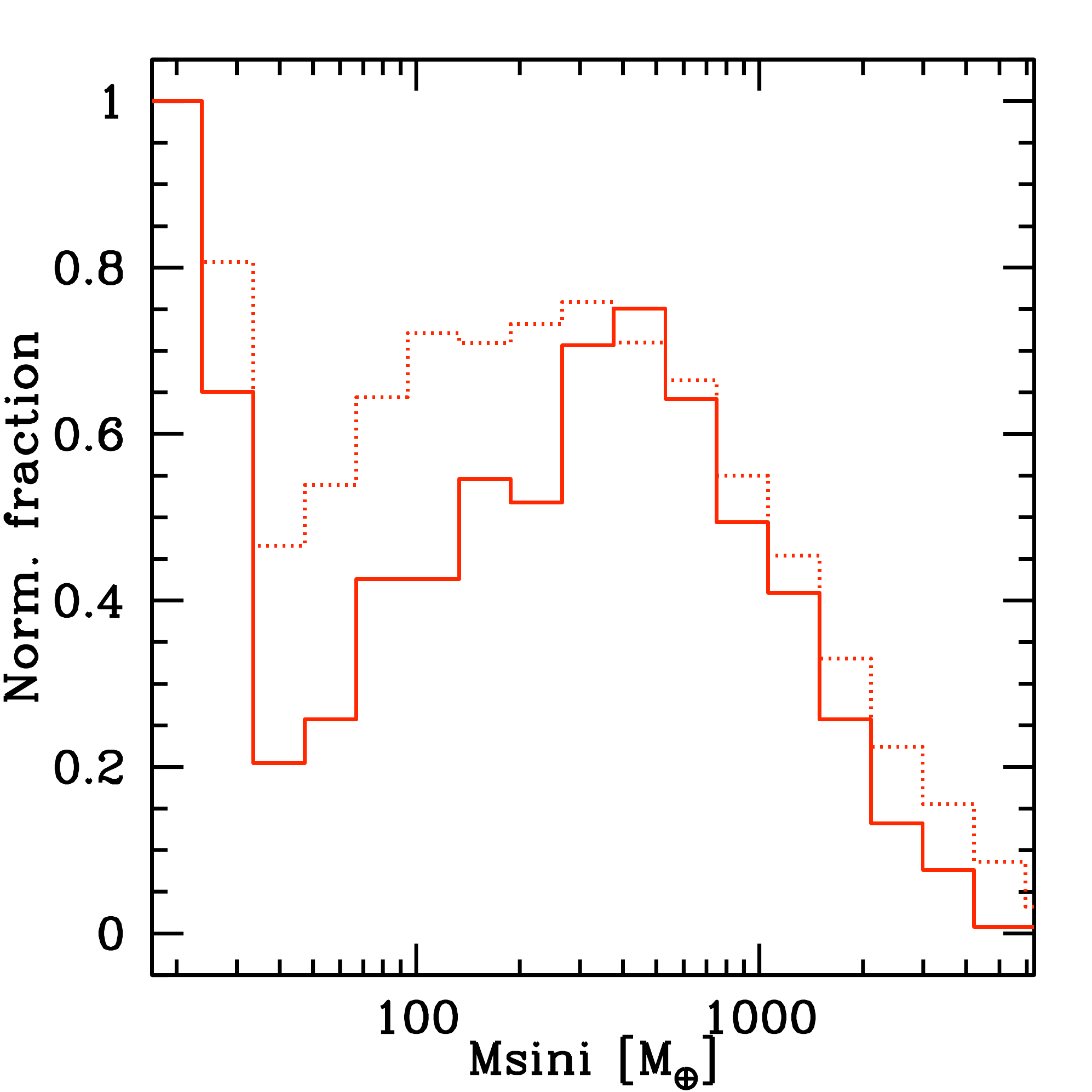}
\caption{Theoretical mass distribution from Neptunian to Jovian mass planets obtained from population synthesis calculations. The solid line shows a population where the planetary gas accretion rate was only limited by the disk accretion rate if the planet has a mass larger than the local gas isolation mass. For the dotted line, the limit was used in any case. Both distributions were normalized to unity at the first bin at about 20 Earth masses. }\label{fig:msiniinter}
\end{figure}

The reason for the difference between the latter two models was discussed in details in Mordasini et al. (\cite{mordasinietal2009a}). It is related to different assumptions about the amount of gas present in the vicinity of the planet as discussed next. Here we illustrate this in Fig. \ref{fig:msiniinter} that shows the theoretically obtained mass distribution from Neptunian to Jovian planets using two different assumptions for the gas accretion rate in runaway. Other settings are similar as in Mordasini et al. (\cite{mordasinietal2009a}). Only planets with a period less than 5 years are shown which are detectable for a RV instrument of 1 m/s precision, similar as {\footnotesize HARPS}. 

In one simulation (solid line), the gas accretion rate of the planet is limited by the accretion rate in the disk only if the mass of the planet is larger than the local gas isolation mass, calculated with the undisturbed gas surface density.  This is similar to the criterion used by Ida \& Lin (\cite{idalin2004}). It means that for masses between roughly 30 up to $\sim100\, \mearth$ i.e. before the gas isolation mass is reached, high gas accretion rates of up to roughly $10^{-2}\,\mearth$/yr can occur, so that the transformation from a Neptunian to a Jovian planet only takes of order $10^{4}$ years, much less than typical disk lifetimes.  The underlying assumption is here that gas already inside the planet's Hill sphere can be accreted independently of the inflow from further away in the disk. This inflow is in turn limited by the disk viscosity. 

In the other simulation (dotted line), the planetary gas accretion rate is limited by the accretion rate in the disk as soon as runaway starts. As cores reach a mass large enough to trigger gas runaway accretion at a moment typically not much before the disk goes away (it is recalled that around about 90\% of FGK stars, there are no giant planets), the accretion rate in the disk has usually already fallen to quite low values at this moment, of order a few $10^{-4}$ to $10^{-3}\,\mearth$/yr.   This means that the transformation from a Neptunian to a Jovian planet takes now several $10^{5}$ years, which is no more much shorter than the remaining disk life time. The underlying assumption is here that due to gap formation, the mass directly available to the planet is in fact small, in particular smaller than the gas isolation mass as calculated above, as (beginning) gap formation reduces the gas surface density around the planet.  This corresponds to the  setting in Mordasini et al. (\cite{mordasinietal2009a}). 

The plot shows that the two different settings have an important effect on the frequency of intermediate mass planets. They are in the first model clearly less frequent that in the second one, with a factor of two of difference for $M\approx40\, \mearth$. Thus, the frequency of intermediate mass planets can be used to directly improve our theoretical understanding of the accretion process.

Bodies with masses of  up to several ten Earth masses can in principle also form after the dissipation of the gas disk if enough solids are available in situ, which is in particular the case at larger distances beyond the ice line. Such bodies would then be essentially gas free. At the rather small orbital distances  of HD\,85390\,b and HD\,103197\,b (1.5 and 0.25  AU, respectively), for realistic solid disk masses, such a formation scenario seems however unlikely: The two planets are larger than what can be formed by this process at their current distances (Ida \& Lin \cite{idalin2004}), considering the host star metallicities.  

The two intermediate mass planets were therefore very probably formed while the gas disk was still present. In this case, the mass of a planet core cannot grow to arbitrarily large values without runaway gas accretion setting in. The real masses of  HD\,85390\,b and HD\,103197\,b which are on the statistical mean a factor $4/\pi$ larger than the projected ones are probably larger than the mass at which this process starts (although not by a large factor, especially for HD\,103197\,b, and the specific value depends on e.g. the unknown core accretion rate, see Papaloizou \& Terquem \cite{papaloizouterquem1999}). The two planets are therefore probably examples where gas runaway started, but only shortly before the gas disk disappeared, so that only low quantities of gas were still available to accrete, and that the gas accretion rate was low. Indeed consist synthetic planets (Mordasini et al. \cite{mordasinietal2009b}) that are situated at a similar position in the mass-distance plane as  HD\,85390\,b ($\msini=42\,\mearth$) of typically about 40\% hydrogen and helium in mass. The scatter around this value is quite large, reflecting different disk properties, and ranges from still clearly solid dominated planets ($\sim10\%$ gas) to small gas giants ($\sim70\%$ gas). For the less massive HD\,103197\,b ($\msini=31.2\,\mearth$) the typical value is about 30\% of gas, with a scatter around this value of about 15\%. We conclude that these two exoplanets have thus probably not only a mass, but also a composition between Neptunian and Jovian planets.

From this  discussion we see that it is important to observationally infer relative frequencies of planetary types between 20 to 100 $\mearth$ to better understand the runaway phase. This will finally enable us to construct better formation models. In the ideal case, the planets should be transiting their host star, but be located still so far away from it that significant evaporation can be ruled out, so that the primordial composition can be studied.

\begin{acknowledgements}
We thank the different observers from other {\footnotesize HARPS} {\footnotesize GTO} subprograms who have also measured the stars presented here, and the observers of HD\,85390 at {\footnotesize CORALIE}. We thank Xavier Dumusque for helpful input. We thank the Swiss National Research Foundation (FNRS) for its continuous support. Christoph Mordasini acknowledges the financial support as a fellow of the Alexander von Humboldt foundation. Nuno C. Santos would like to thank the support by the European Research Council/European Community under the FP7 through a Starting Grant, as well from Funda\c{c}\~ao para a Ci\^encia e a Tecnologia (FCT), Portugal, through programme Ci\^encia\,2007, and in the form of grants reference PTDC/CTE-AST/098528/2008 and PTDC/CTE-AST/098604/2008. We thank an anonymous referee for helpful comments.

\end{acknowledgements}

\end{document}